\documentclass[aps,twocolumn,superscriptaddress,showpacs]{revtex4}
\usepackage{epsf,latexsym}
\usepackage{times}

\newcommand{\bra}[1]{\langle #1 |}
\newcommand{\ket}[1]{| #1 \rangle}

\newcommand{\be}{\begin{equation}}
\newcommand{\ee}{\end{equation}}

\newcommand{\ignore}[1]{}

\def\CC{{\rm\kern.24em \vrule width.04em height1.46ex depth-.07ex
    \kern-.30em C}}
\def\P{{\rm I\kern-.25em P}}

\def\RR{{\rm
         \vrule width.04em height1.58ex depth-.0ex
         \kern-.04em R}}

\def\bbbone{{\mathchoice {\rm 1\mskip-4mu l} {\rm 1\mskip-4mu l}
{\rm 1\mskip-4.5mu l} {\rm 1\mskip-5mu l}}}

\def\bbbc{{\mathchoice {\setbox0=\hbox{$\displaystyle\rm C$}\hbox{\hbox
to0pt{\kern0.4\wd0\vrule height0.9\ht0\hss}\box0}}
{\setbox0=\hbox{$\textstyle\rm C$}\hbox{\hbox
to0pt{\kern0.4\wd0\vrule height0.9\ht0\hss}\box0}}
{\setbox0=\hbox{$\scriptstyle\rm C$}\hbox{\hbox
to0pt{\kern0.4\wd0\vrule height0.9\ht0\hss}\box0}}
{\setbox0=\hbox{$\scriptscriptstyle\rm C$}\hbox{\hbox
to0pt{\kern0.4\wd0\vrule height0.9\ht0\hss}\box0}}}}

\def\bbbz{{\mathchoice {\hbox{$\sf\textstyle Z\kern-0.4em Z$}}
{\hbox{$\sf\textstyle Z\kern-0.4em Z$}}
{\hbox{$\sf\scriptstyle Z\kern-0.3em Z$}}
{\hbox{$\sf\scriptscriptstyle Z\kern-0.2em Z$}}}}

\newcommand{\putfig}[2]{$$\leavevmode\hbox{\epsfxsize=#2 cm
   \epsffile{#1.eps}}$$}

\begin{document}
\title{String and Membrane condensation on 3D lattices}
\author{Alioscia Hamma}
\affiliation{Institute for Scientific Interchange (ISI), Villa
Gualino, Viale Settimio Severo 65, I-10133 Torino, Italy}
\affiliation{Dipartimento di Scienze Fisiche, Universit\`a
Federico II, Via Cintia ed. G, 80126, Napoli, Italy}

\author{Paolo Zanardi}\affiliation
{Institute for Scientific Interchange (ISI), Villa
Gualino, Viale Settimio Severo 65, I-10133 Torino, Italy}
\author{Xiao-Gang Wen}
\homepage{http://dao.mit.edu/~wen}
\affiliation{Department of Physics, Massachusetts Institute of Technology,
Cambridge, Massachusetts 02139}
\begin{abstract}
In this paper, we investigate the general properties of lattice spin models
that have string and/or membrane condensed ground states.  We discuss the
properties needed to define a string or membrane operator.  We study three 3D
spin models which lead to $Z_2$ gauge theory at low energies.  All the three
models are exactly soluble and produce topologically ordered ground states.
The first model contains both closed-string and closed-membrane condensations.
The second model contains closed-string condensation only. The ends of
open-strings behave like fermionic particles. The third model also has
condensations of closed membranes and closed strings.  The ends of open
strings are bosonic while the edges of open membranes are fermionic.  The
third model contains a new type of topological order.
\end{abstract}
\maketitle

\section{Introduction}

The discovery of Fractional Quantum Hall liquids (FQH liquids) by Tsui,
Stormer and Gossard \cite{tsui} in 1982 showed that not all the states of
matter are associated with symmetries (or the breaking of symmetries). As an
example, FQH liquids cannot be described by the Landau's theory
\cite{landau,ginzburg} of broken symmetries and local order
parameters\cite{wenorder,Wbook}.  In Landau's theory,
the internal order is defined by the symmetry of the states. Symmetry is an
universal property of the states, i.e.  a property shared by all the states in
the same phase. The symmetry group (SG) can thus characterize the internal
orders of those states.  The internal order of FQH liquids is the internal
structure of their quantum ground state.  The kind of order that explains this
internal structure of FQH liquids is a {\em topological order}
\cite{wenorder,Wbook}. Topological order is a special case of more general
quantum order \cite{wenqorder,Wbook}. Topological/quantum order cannot be
characterized by symmetry breaking since all FQH states have the same
symmetry.  To characterize quantum orders, we need to find universal
properties of the wave function. One way to characterize quantum orders is
through the projective symmetry group (PSG) \cite{Wpsg}. PSG is the group of
symmetry of the mean-field ansatz of a mean-field Hamiltonian $H_{mean}$ that
describes a quantum ordered state.  Two different physical wave functions
obtained from two different mean-field ansatz can have the same symmetry and
{\em different} PSG. Thus the PSG gives a more refined characterization of the
internal orders than the SG and can describe those internal orders that are
not distinguished by the symmetry group.

The topological order, as a special case of quantum order, is a quantum order
where all the excitations above ground states have finite energy gaps. FQH
liquids present non trivial topological orders in which the degeneracy of the
ground state depends on the topology of the space
\cite{degeneracy,zee}. The ground state of FQH liquids on a Riemann
surface of genus $g$ is $\tilde{q}^g-$fold degenerate where $\tilde{q}$ is the
ground state degeneracy in a torus topology. This degeneracy is robust against
any perturbations. For a finite system of size $L$ the ground state degeneracy
is lifted and the energy splitting is $\sim e^{-L/\xi}.$ This robustness is at
the root of the proposal of fault tolerant quantum computation at the physical
level \cite{kitaev}.

A particular class of quantum orders is the one from string condensation
\cite{lightorigin}. We say we have string condensation in the
ground state $\ket{\xi}$ when\\
(a) certain closed-string operator,
$W(\overline{\gamma})$, satisfies
\begin{equation}
\bra{\xi}W(\overline{\gamma})\ket{\xi}=1.
\end{equation}
(b) the closed-string operator cannot be decomposed in smaller pieces 
$
W(\overline{\gamma})
=W(\gamma_1) W(\gamma_2) \cdots W(\gamma_n) 
$
where each piece satisfies 
$\bra{\xi}W(\gamma_l)\ket{\xi}=1.$

A string-net is a branched string.
It turns out \cite{lightorigin} that quantum ordered states that
produce and protect massless gauge bosons and massless fermions are string-net
condensed states. Moreover, different string-net condensations are not
characterized by symmetries, but by projective symmetry group. In this case,
PSG describes the symmetry in the hopping Hamiltonian for the ends of
condensed strings.  Then the characterization of different string-net
condensations classifies different topological/quantum orders. Systems that
feature string-net condensation, if gapped, feature a ground state degeneracy
that depends on the topology of the system and that is robust against
arbitrary local perturbations.  

Ends of open strings are particle-like objects which can have a non trivial
statistics \cite{levin}.  When on a two-dimensional system such a quasi
particle winds around another quasi-particle of a different kind, its wave
function picks a phase. The particle has undergone an Aharonov-Bohm effect,
whose topological nature is described by a Cherns-Simons theory \cite{bais}.
This phenomenon corresponds to the fact that the end of an open string 
can be detected by by a closed string of another type that encloses its end.

On a three dimensional lattice, the end of an open string can be detected by a
closed surface. So we can inquire the meaning of the condensation of closed
membranes.
Similar to closed-string condensation, closed-membrane condensation is a
superposition of closed membranes of arbitrary sizes, shapes, and numbers.
Just like closed-string condensation, closed-membrane condensation also
implies topological order. Many 3D models have both closed-string and
closed-membrane condensation due to a natural duality between string and
membrane in 3D space.  But it may be possible to have 3D models with only
string condensation.  We will present an example.

We also build a new model what a new kind of closed string and membrane
condensation. We argue that this model has new type of topological order.

\section{$Z_2$ lattice gauge theory -- a model with 
string and membrane condensation} \label{membranes}

Let us consider a three dimensional cubic lattice. Then we can place a
spin-1/2 on
each link of the lattice. 
%We can define strings, i.e. operators defined on
%curves and membranes, i.e.operators defined on surfaces. 
A string operator can
be defined by drawing a curve $\gamma$ connecting the sites of the lattice and
acting with a $\sigma^z$ on all the links belonging to $\gamma$: 
\begin{equation}
W[\gamma]=\prod_{j \in \gamma}\sigma^z_{j}.
\end{equation}
A membrane operator $M[\Sigma]$ is obtained by drawing a two-dimensional
surface $\Sigma$ in the dual lattice and acting with a spin flip $\sigma^x$ on
all the links orthogonal to $\Sigma$:
\begin{equation}
M[\Sigma]=\prod_{j\bot\Sigma}\sigma^x_j.
\end{equation}
As expected, closed membranes are able to detect the ends of open strings.
The open string flips the spin on the membrane where it punctures it because
the membrane operator anti-commutes with the string operator when they
intersect in only one point,
\begin{equation}
\label{strmemalg}
W[\gamma]M[\overline{\Sigma}]\ket{\psi}
=-M[\overline{\Sigma}]W[\gamma]\ket{\psi},
\end{equation} 
where $\overline{\Sigma}$ is a closed surface in the dual lattice.  If the sign
of a closed membrane-state is flipped we know that there is the end of an open
string inside. If the string punctures the closed membrane in two points, then
the end of the string is not inside and the sign of the state will not be
flipped because the two operators would in fact commute. A closed membrane can
detect the presence of a particle (the end of an open string) inside even if
the membrane is actually very far from the particle.

We would like to stress that not all products of operators along a string give us a non-trivial string operator. Similarly, not all products of operators on a
membrane give us a non-trivial membrane operator.  The product of identity
operators along a string or on a membrane is an example of trivial
string-operator or membrane operator.  However, in our case, the non-trivial
algebraic relation (\ref{strmemalg}) between the large string-operator and
membrane operator ensures that both the string-operator and the membrane-operator defined above are non-trivial.

A plaquette operator is the product of $\sigma^z$ on all the spins belonging
to a same plaquette $p$.  A closed string operator $W[\overline{\gamma}]$ can
also be expressed as a product of plaquette operators.  When we multiply two
neighbouring plaquette operators, the $\sigma^z$ acts twice on the shared
link, and so the resulting operator is the product of $\sigma^z$ on the border
of the two plaquettes. 

A star operator is
on the other hand the product of $\sigma^x$ on all the links extruding from a
site $s$: $A_s=\prod_{j\in s}\sigma^x_j$. Similarly then, a closed membrane
operator $M[\overline{\Sigma}]$ is the product of all the star operators
enclosed in such a two-dimensional closed surface $\overline{\Sigma}$:
\begin{equation}
M[\overline{\Sigma}]
=\prod_{j\bot \overline{\Sigma}}\sigma^z_j=\prod_{s\in V}A_s,
\end{equation}
where $V$ is the volume enclosed in $\overline{\Sigma}$:
$\overline{\Sigma}=\partial V.$ The star operator is then the operator
corresponding to the elementary closed membrane, the cube. If we consider in
the dual lattice the six faces orthogonal to the links of a star, we see that
they form a cube. Since when multiplied with each other the stars cancel their
interior, they are surface operators and not volume operators. So in general,
the product of two membrane operators is still a membrane operator because the
interior cancels.

An example of a model featuring both membrane and string condensation is
given by the following Hamiltonian:
\begin{equation}\label{spin}
H_{spin}=-g \sum_{p}\prod_{j\in p}\sigma^z_j -U\sum_{s}\prod_{j\in s}\sigma^x_j,
\end{equation}
so that the Hamiltonian is the sum of all the plaquette operators and star
operators. Such a Hamiltonian defines a $Z_2$ lattice gauge theory
\cite{wegner}.  The model is exactly soluble since all the plaquette operators
and star operators commute with each other \cite{kitaev}.  

Having a spin on each link, the dimension of the local Hilbert space is $2.$
If $N$ is the number of the sites, on a cubic lattice we have $3N$ links
\footnote{On a cubic lattice and periodic boundary conditions with $N$ sites there are $N$ cubes and $3N$ links, plaquettes, and faces.}. The dimension of the global Hilbert space is
hence $2^{3N}.$ How many states can we label with the operators $\prod_{j\in
p}\sigma^z_j$ and $\prod_{j\in site}\sigma^x_j$? We have $N$ star operators,
and $3N$ plaquettes in $3D$. However, not all these plaquettes are
independent. Indeed, in each cube the product of the eight plaquettes is
identically one, as it is immediate to verify. This gives us $N$ constraints
on the plaquettes. The number of independent plaquettes is thus $3N-N=2N.$
Together with the star operators, we can then label all the $2^{3N}$ states.
We have a finite degeneracy of the ground state due to topological global
constraints on the star and plaquette operators (if we are on a torus for
instance).  

This model features both closed-string and closed-membrane condensation.
Closed-string operators and closed-membrane operators both commute with the
Hamiltonian because every closed string (membrane) shares either $0$ or $2$
links with any star (plaquette):
\begin{eqnarray}
\left[H_{spin},W[\overline{\gamma}]\right]=0\\
\left[H_{spin},M[\overline{\Sigma}]\right]=0
\end{eqnarray} 
Thus the ground state is an eigenstate of $W[\overline{\gamma}]$
and $M[\overline{\Sigma}]$ with eigenvalue $\pm 1$ since
$W[\overline{\gamma}]^2=M[\overline{\Sigma}]^2=1$.
This leads to a condensation of
the closed-string and the closed-membrane membrane operators
$<W[\overline{\gamma}]>=<M[\overline{\Sigma}]>=\pm 1$ regardless the size
of the strings and the membranes.

Physically, when acting on the ground state, the open-string operator creates
a pair of $Z_2$ charges at its ends, while the open-membrane operator creates
a loop of $Z_2$ flux at it edge. So it is natural that a 3D $Z_2$ gauge theory
has both the closed-string and the closed-membrane condensations.

In order to have string or membrane condensation, it is important that they do
not dissolve, i.e. they are not decomposable into smaller objects that also
condense. 
For example, a  closed-string operator can be written as
a product of two open string operators
$ W[\overline{\gamma}] =W[\overline{\gamma_1}] W[\overline{\gamma_2}] $.
We require that the open string operators do not condense. That is
\begin{equation}
 \frac{<W[\overline{\gamma_1}]>< W[\overline{\gamma_2}]>}
{<W[\overline{\gamma}]>} \to 0
\end{equation}
as the size of strings approach infinity. If closed
membranes can be written as the product of smaller objects, and each smaller
object still condenses, then there is no need to talk about membrane
condensation, and we cannot expect having topological order. It is the fact
that a big closed membrane -that can explore the topology of the lattice-
condenses that implies a topological order. For the same reason open membranes
must be forbidden in the ground state. We can obtain it by making them paying
an energy or by means of a constraint, as we will see in section
\ref{omembranes}. 

We would like to remark that the theory (\ref{spin}) can be mapped into a
model with Majorana fermions on the links. We put six Majorana fermions at
each site and one so called 'ghost' Majorana fermion at each link. Then we
"move" the Majorana fermions from the sites to the midpoints of each link
according to the directions shown in Fig.\ref{h3d}. So at this point we have
three Majorana fermions at each link, two coming from the sites plus the
initial ghost Majorana fermion. The mapping is given by the following
representation of the Pauli matrices: 
\begin{equation}\label{ghostmap}
\sigma^i=\epsilon^i_{jk}\lambda^j\lambda^k.  
\end{equation}

\section{A model with string condensation only}\label{stringsonly}

\begin{figure}
\putfig{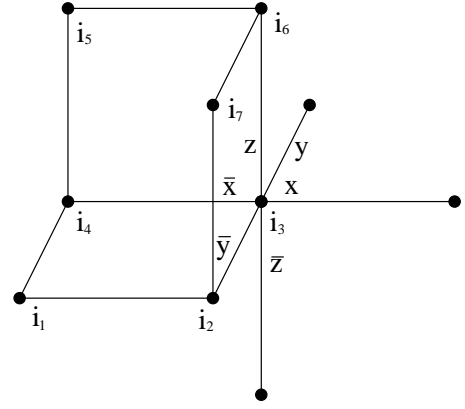}{6}
\caption{A three dimensional model with Majorana fermions on the sites. The
six Majorana fermions label the six vectors from a site in the following way:
$x\mapsto +\hat{x}, \overline{x}\mapsto -\hat{x},y\mapsto \hat{x},
\overline{y}\mapsto -\hat{y},z\mapsto \hat{z}, \overline{z}\mapsto -\hat{z}.$
The plaquettes in the three planes are shown.
$\hat{F}_{p_{xy}},\hat{F}_{p_{yz}},\hat{F}_{p_{zx}}$ are respectively the
plaquettes $\hat{F}_{{\bf i}_1{\bf i}_2{\bf i}_3{\bf i}_4},\hat{F}_{{\bf
i}_2{\bf i}_3{\bf i}_6{\bf i}_7},\hat{F}_{{\bf i}_4{\bf i}_3{\bf i}_6{\bf
i}_5}$ where ${\bf i}_1={\bf i};{\bf i}_2={\bf i+\hat{x}};{\bf i}_3={\bf
i+\hat{x}+\hat{y}};{\bf i}_4={\bf i+\hat{y}};{\bf i}_5={\bf
i+\hat{y}+\hat{z}};{\bf i}_6={\bf i+\hat{x}+\hat{y}+\hat{z}};{\bf i}_7={\bf
i+\hat{x}+\hat{z}}.$} \label{h3d} 
\end{figure}

 In this section we want to show a model that has closed string condensation
but no closed-membrane condensation. Membranes either pay energy or,
if they commute with the Hamiltonian, dissolve in smaller pieces.

 Let us consider the following exactly solvable model on a cubic lattice
\cite{lightorigin,levin}. We introduce six Majorana fermions at each
site of the lattice, namely $\lambda_{\bf i}^a,$ where
$a=x,\overline{x},y,\overline{y},z,\overline{z}.$  Define the plaquette
operator in the plane $\alpha\beta$:
\begin{equation}
\hat{F}_{p_{\alpha\beta}}=-\lambda_{{\bf i}}^{\beta}\lambda_{{\bf i}}^{\alpha}
\lambda_{{\bf i+\hat{\alpha}}}^{\overline{\alpha}}\lambda_{{\bf
i+\hat{\alpha}}}^{\beta}\lambda_{{\bf
i+\hat{\alpha}+\hat{\beta}}}^{\overline{\beta}}\lambda_{{\bf
i+\hat{\alpha}+\hat{\beta}}}^{\overline{\alpha}}\lambda_{{\bf
i+\hat{\beta}}}^{\alpha}\lambda_{{\bf i+\hat{\beta}}}^{\overline{\beta}},
\end{equation}
with $\alpha,\beta=x,y,z$
The Hamiltonian is then
\begin{eqnarray}\label{wen3D}
%\nonumber
 & &H_{3D}=
\ignore{
-g\sum_{\Box}(\lambda_{{\bf i}}^y\lambda_{{\bf i}}^x
\lambda_{{\bf i+\hat{x}}}^{\overline{x}}\lambda_{{\bf i+\hat{x}}}^y
\lambda_{{\bf i+\hat{x}+\hat{y}}}^{\overline{y}}\lambda_{{\bf
i+\hat{x}+\hat{y}}}^{\overline{x}}
\lambda_{{\bf i+\hat{y}}}^x\lambda_{{\bf i+\hat{y}}}^{\overline{y}} +
\\
\nonumber
& &\lambda_{{\bf i+\hat{x}}}^z\lambda_{{\bf i+\hat{x}}}^y
\lambda_{{\bf i+\hat{x}+\hat{y}}}^{\overline{y}}\lambda_{{\bf
i}+\hat{x}+\hat{y}}^z
\lambda_{{\bf i+\hat{x}+\hat{y}+\hat{z}}}^{\overline{z}}
\lambda_{{\bf i+\hat{x}+\hat{y}+\hat{z}}}^{\overline{y}}
\lambda_{{\bf i+\hat{x}+\hat{z}}}^y\lambda_{{\bf
i+\hat{x}+\hat{z}}}^{\overline{z}}+
\\
\nonumber
& & \lambda_{{\bf i+\hat{y}}}^z\lambda_{{\bf i+\hat{y}}}^y
\lambda_{{\bf i+\hat{x}+\hat{y}}}^{\overline{y}}\lambda_{{\bf
i+\hat{x}+\hat{y}}}^z
\lambda_{{\bf i+\hat{x}+\hat{y}+\hat{z}}}^{\overline{z}}\lambda_{{\bf
i+\hat{x}+\hat{y}+\hat{z}}}^{\overline{y}}
\lambda_{{\bf i+\hat{y}+\hat{z}}}^y\lambda_{{\bf
i+\hat{y}+\hat{z}}}^{\overline{z}})
\\
}
g\sum_{p}(\hat{F}_{{p}_{xy}}+\hat{F}_{{p}_{yz}}+\hat{F}_{{p}_{zx}}),
\end{eqnarray}
where the sum is taken on all the plaquettes in the $xy,yz,zx$ planes (see
Fig.\ref{h3d}).

 This model is exactly solvable because all the plaquette operators commute
with each other $[\hat{F}_{{\Box}_i},\hat{F}_{{\Box}_j}]=0.$

 Let us introduce now the following complex fermion operators at each site
${\bf i}$.
\begin{equation}
2\psi_x=\lambda^x
+i\lambda^{\overline{x}},2\psi_y=\lambda^y
+i\lambda^{\overline{y}},2\psi_z=\lambda^z
+i\lambda^{\overline{z}}.
\end{equation}

 We project down to the physical Hilbert space with an even number of fermions
per site ${\bf i}$:
\begin{equation}\label{evenfermions}
(-)^{\sum_{a=x,y,z}\psi^{\dagger}_{a,{\bf i}}\psi_{a,{\bf i}}}=1.
\end{equation}
 The projection above makes the theory a gauge theory. The physical
states are invariant under local $Z_2$ transformations generated by
\begin{equation}\label{gauge3d}
\hat{G}=\prod_{\bf i}G_{\bf i}^{\sum_{a=x,y,z}\psi^{\dagger}_{a,{\bf i}}\psi_{a,{\bf i}}},
\end{equation}
where $G_{\bf i}$ is an arbitrary function on the sites {\bf i} with the only
two values $\pm1.$

The Hamiltonian (\ref{wen3D}) acts on spin $3/2$ states
\cite{levin,lightorigin} by means of the following mapping:
\begin{equation}\label{gamma}
\gamma^{ab}_{\bf i}=\frac{i}{2}(\lambda^a_{\bf i}\lambda^b_{\bf i}-\lambda^b_{\bf i}\lambda^a_{\bf i}).
\end{equation}
The operators $\gamma^{ab}_{\bf i}$ act on the local $4$-dimensional physical
Hilbert space, that is, after the projection. In terms of the
$\gamma^{ab}_{\bf i}$ we can write down the Hamiltonian acting on spin-$3/2$
states \cite{lightorigin}:
\begin{equation}
H_{3D}\mapsto H_{3/2}=
g\sum_{p}(\gamma^{ab}_{\bf i}\gamma^{\overline{b}a}_{\bf i+\hat{b}}\gamma^{\overline{a}\overline{b}}_{\bf i+\hat{b}+\hat{a}}\gamma^{b\overline{a}}_{\bf i+\hat{a}}).
\end{equation}
 The ground state of $H_{3D}$ has closed-string condensation. This means that
we can define closed strings that commute with the Hamiltonian $H_{3D}.$ What
kind of strings can we define in this model (or similar models)? We can define
two types of strings running on links: strings that end at the midpoint of the
links and strings that end on the sites. The open string operator of the
former type is
\begin{equation}\label{stringI}
W[\gamma]_{I}=-
(\lambda^{a_1}_{{\bf i}_1}\lambda^{b_1}_{{\bf i}_1})
(\lambda^{a_2}_{{\bf i}_2}\lambda^{b_2}_{{\bf i}_2})
...
(\lambda^{a_n}_{{\bf i}_n}\lambda^{b_n}_{{\bf i}_n}),
\end{equation}
where $\gamma$ is an open string running on the lattice links. Strings that end on sites would decompose in edges all commuting with the Hamiltonian and hence not giving a {\em closed} string condensation, therefore we will focus only on the strings of the type Eq.(\ref{stringI}). The pair of
indices $ab$ gives the shape of any element of the string. $\lambda^{a}_{{\bf
i}}\lambda^{b}_{{\bf i}} =\lambda^{x}_{{\bf i}}\lambda^{\overline{x}}_{{\bf
i}} $ is associated with an horizontal string straddling the site ${\bf i},$
while $\lambda^{\overline{y}}_{{\bf i}}\lambda^{x}_{{\bf i}}$ is associated
with a L-shaped elementary string crossing the site ${\bf i}.$ These strings
have endpoints sitting at the midpoint of links.   Moreover, if we
close a string around the elementary loop (the square), we obtain the
plaquette operator:
\begin{equation}
W[\gamma_{p}]=\hat{F}_{p}.
\end{equation}

Large (contractible) loops are products of this elementary loops:
$W[\partial\Sigma]=\prod_{p\in\Sigma}\hat{F}_p,$ where $\Sigma$ is a
surface made of squares of the lattice and $\partial\Sigma$ its contour. We
notice that the product of loops with some overlap gives a loop and not a net because the interior cancels. In fact the interior is identically one because the Majorana fermions square the identity. The important fact is that loops -contractible or not- commute with the Hamiltonian, as it is easy to check:
\begin{equation}
[H_{3D},W[\overline{\gamma}]]=0.
\end{equation}
So the ground state of $H_{3D}$ has closed-string condensation.

 We want now to argue about the degeneracy of the ground state if the lattice
has a torus topology. The dimension of the total Hilbert space is computed in
the following way. We haves six Majorana fermions per site and with them we
defined three complex fermion operators. Three complex fermion operators
generate a eight-dimensional local Hilbert space at each site. So the
dimension of the total Hilbert space is $8^N.$ The projection onto the
physical Hilbert space at each site gives us thus a $4$-dimensional local
Hilbert space at each site, because there are four states out of eight with
even number of fermions. Therefore after the projection the Hilbert space is
$4^N=2^{2N}-$dimensional. How many states can we label with the commuting
operators $\hat{F}_{p}$? We have $3N$ such operators, but not all of them
are independent. We have local constraints and global constraint on them. The
local constraint is given by the fact -which is immediate to prove- that in
each cube the product of all the plaquettes is identically equal to one:
\begin{equation}
\prod_{p\in c}\hat{F}_{p}=1,
\end{equation}
where $c$ labels the cubes in the lattice. This gives us $N$ constraints. The
global constraints are given by the periodic boundary conditions. These
conditions provide $3$ constraints. It turns out that the number of
independent plaquette operators $\hat{F}_{p}$ is $3N-N-3=2N-3.$ So we can
label $2N-3$ states out of $3N$ and this means we are left with a
$2^{2N-(2N-3)}=8$-fold degeneracy of the ground state. This degeneracy is due
to the closed loop condensation. 

Notice that closed strings are not decomposable in smaller objects that still
commute with the Hamiltonian. Small elements -called 'dimers'- of the type
$\lambda_{\bf i}^{a}\lambda_{\bf i}^b$ (which is an elementary string of the I
type) never commute with some of the plaquettes $\hat{F}_{p}$.

\begin{figure}
\putfig{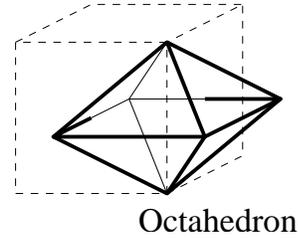}{4}
\caption{
The dash-lines represent the cube in the cubic lattice.
The solid lines represent the octahedron.
} \label{oct}
\end{figure}

What we want to argue now is that the model Eq.(\ref{wen3D}) is a model with
string condensation only.  The membrane operator that we want to construct
should trap the ends of strings which now live on the centers of the links.
Because of this, it is not natural to use the faces of the cubic lattice or
the dual lattice to form the membrane.  The dual lattice of the links is
formed by the octahedron's (see Fig.  \ref{oct}). We can put many octahedrons
together to form a volume.  The surface of such a volume is a natural choice of
membrane which traps the centers of the links.  The elementary (the smallest)
membrane corresponds to the faces of a single octahedron.  What is the
operator for the elementary membrane?  Notice that each elementary membrane
contain a single link, say $<{\bf i, i+\hat{x}}>$.  So a natural choice of the
elementary membrane operator is $\lambda_{\bf i}^x\lambda_{\bf i +
\hat{x}}^{\overline{x}}$.  A generic membrane operator is the product of the
elementary membrane operators for the enclosed octahedrons.
%There are no possible membranes. 
On each interior lattice site, ${\bf j}$, enclosed by the membrane, we have a
product
\begin{equation}
\prod_{\alpha\in{\bf j}}\lambda_{\bf j}^{\alpha}
\end{equation}
where $\alpha=x,\overline{x},y,\overline{y},z,\overline{z}$.  Since in the
projected physical Hilbert space, $\prod_{\alpha\in{\bf j}}\lambda_{\bf
j}^{\alpha}=1$.\cite{lightorigin} So the product of the elementary membrane
operators is a membrane operator which only acts on the sites on the membrane.
The membrane operator also satisfies an important condition that it commutes
with all the closed-string operators. The membrane operator anti-commutes with
the open-string operators of one end of the open string is enclosed by the
membrane. So the membrane operator can detect the presence of the trapped ends
of strings.  However, the membrane operator, in general, changes the fermion
number by an odd number (on a site on the membrane).  So the membrane operator
defined above, although having many of the right properties, does not act
within the physical Hilbert space.

Let us compare the model  $H_{3D}$ and the model Eq.(\ref{spin}).  Notice that
the plaquette terms in $H_{3D}$ map well onto the spin model Eq.(\ref{spin}).
We can associate $\sigma^z$ with $i\lambda^{\overline{a}}_{\bf
i}\lambda^a_{\bf i}$ and see that $\hat{F}_{p}\mapsto \prod_{j\in
p}\sigma^z_j.$ So the string operator in the  model  $H_{3D}$ can be mapped
into the string operator in the model Eq.(\ref{spin}).  What does not map well
is the star term for the reasons stated above. In order to build a good star
operator so that the star maps onto the term $-U\sum_{s}\prod_{j\in
s}\sigma^x_j,$ we need to put an additional 'ghost' Majorana fermion on each
link and then realize the mapping Eq.(\ref{ghostmap}).  Because of this, we
obtain the membrane operator in  the  model  $H_{3D}$ from the membrane
operator in the model Eq.(\ref{spin}).

After many trials, we fail to obtain a membrane operator with the right
properties. This leads us to believe that the model $H_{3D}$
(\ref{wen3D}) has no membrane
condensation.

\section {Exactly solvable model with membrane condensation}\label{omembranes}
The two models discussed above were constructed to have string condensation.
The first model also has a membrane condensation.
In this section we will directly construct a model
that has a membrane condensation. The model is the following. We
start with a cubic lattice with $N$ sites and put four Majorana fermions at
each link, thus we have $12N$ Majorana fermions in total.  We label the
Majorana fermions according to the directions orthogonal to the link. For
example, on a link $<{\bf i,i+\hat{x}}>$ we have the following Majorana
fermions:
\be
\lambda^y_{<{\bf i,i+\hat{x}}>},\ \lambda^{\overline{y}}_{<{\bf
i,i+\hat{x}}>},\ \lambda^z_{<{\bf i,i+\hat{x}}>},\
\lambda^{\overline{z}}_{<{\bf i,i+\hat{x}}>}
\ee
We can define the complex fermion operators at each link $<{\bf i,i+\hat{a}}>$:
\begin{equation}
2\psi_{b,<{\bf i,i+\hat{a}}>}=\lambda^b_{<{\bf i,i+\hat{a}}>}+i\lambda^{\overline{b}}_{<{\bf i,i+\hat{a}}>}
\end{equation}
where $a,b=x,y,z$ and $a\ne b.$ Of course on a link ${\bf <i,i+\hat{x}>}$ we
can only define $\psi_y$ and $\psi_z$ and so on.  So at each link we have
defined two complex fermion operators and thus at each link sits a
4-dimensional local Hilbert space $\mathcal{H}$. Since the number of links on a $3D$ cubic
lattice with $N$ sites is $3N$, 
the total Hilbert space $\mathcal{H}^{\otimes 3N}$ has dimension $4^{3N}=2^{6N}.$
\begin{figure}
\putfig{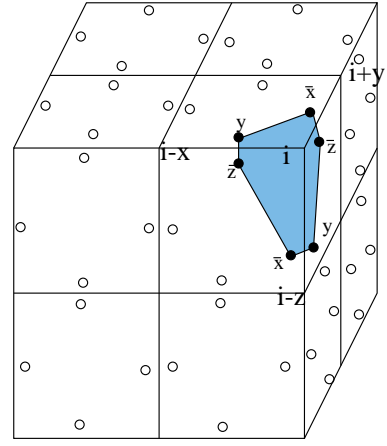}{5}
\caption{The Cube operator and the Corner Loop operator with the nomenclature
of the Majorana fermions forming a corner loop.} \label{cubeoperator}
\end{figure}

On each link we define the Link operator in this way:
\be\label{linkterm}
\hat{L}_{<{\bf i,i+a}>}=\lambda_{<{\bf i,i+a}>}^b\lambda_{<{\bf i,i+a}>}^{\overline {b}} \lambda_{<{\bf i,i+a}>}^c\lambda_{<{\bf i,i+a}>}^{\overline{c}}
\ee
where $a \ne b \ne c$ and they take values in the set $\{x,y,z\}$.
In a cubic lattice each link is shared by four crossing faces and every face has as contour four links. 
For each link, we can uniquely associate 
one Majorana fermion to each of the four faces that share that link. Since
each face is bordered by four links, each face receives a total of four
Majorana fermions from the links that border it. This assignment is univocal.
Each Majorana fermion is assigned to one and only one face. The corresponding face operator is defined as
\be\label{faceoperator}
\hat{F}_{s}=\lambda_{<{\bf i,i+a}>}^b \lambda_{<{\bf i+a,i+a+b}>}^{\overline{a}} \lambda_{<{\bf i+a+b,i+b}>}^{\overline{b}} \lambda_{<{\bf i+b,i}>}^a
\ee
where the face $s$ is on the plane $\{ab\}$ and $a \ne b$ and again they take the values $\{x,y,z\}$.
Notice that the face operator corresponds to a link operator in the dual lattice. Notice also that the operator on the link $<l>$ anti-commutes with any of the four adjacent face operators because they have a Majorana fermion in common:
\be
\{\hat{L}_{<l>},\hat{F}_{s}\}_+=0
\ee
After the discussion of the previous sections, we know that we can define a good elementary closed membrane operator by taking the product of the six face operators on a cube. This is equivalent to a star operator on the dual lattice. The cube operator is then:
\be\label{cube}
\hat{S}_c=\prod_{s\in c}\hat{F}_{s}
\ee
The cube operator shares two Majorana fermions with any adjacent link $<l>$ so they commute:
\be
\left[\hat{L}_{<l>},\hat{S}_c\right]=0
\ee
bigger membrane operators are the product of all the faces that make the surface on which the operator is defined:
\be\label{membraneoperator}
\hat{M}_{\Sigma}=\prod_{s\in\Sigma}\hat{F}_s
\ee
Open membranes do not commute with the link operator, because they share one Majorana fermion on the border of the membrane. But a closed membrane operator, being the product of cubes, does commute with the link operator.

The last operator we want to define is the Corner Loop operator. It is the operator that collects in a loop the six Majorana fermions that are associated to each corner of a cube. for instance the corner made by, say, the links
${\bf<i,i+\hat{x}>,<i+\hat{x},i+\hat{x}+\hat{y}>,<i+\hat{x},i+\hat{x}-\hat{z}}>$
is labeled by the vector ${\bf j}=(\overline{x},y,\overline{z})$ and the
associated Corner Loop operator $\hat{C}^{\bf j}$ is
\begin{eqnarray} \label{corner}
\nonumber
\hat{C}_{\bf i}^{\overline{x}y\overline{z}}&\equiv&
\lambda^{\overline{z}}_{\bf<i,i+\hat{x}>}
\lambda^y_{\bf<i,i+\hat{x}>}
\lambda^{\overline{x}}_{\bf <i+\hat{x},i+\hat{x}+\hat{y}>}
\lambda^{\overline{z}}_{\bf <i+\hat{x},i+\hat{x}+\hat{y}>}
\\
&\times&
\lambda^y_{\bf <i+\hat{x},i+\hat{x}-\hat{z}>}
\lambda^{\overline{x}}_{\bf <i+\hat{x},i+\hat{x}-\hat{z}>}
,
\end{eqnarray}
see Fig.\ref{cubeoperator}.
At each site ${\bf i}$ there are eight corners that are labeled by ${\bf j}=(a,b,c)$ with $a=x,\overline{x}, b=y,\overline{y},c=z,\overline{z}$.

So the expression for the generic Corner Loop operator on the site ${\bf i}$ is
\begin{eqnarray}
\nonumber
\hat{C}^{(a,b,c)}_{\bf i}&=&
\lambda^b_{<{\bf i, i+a}>}\lambda^c_{<{\bf i, i+a}>}
\lambda^a_{<{\bf i, i+b}>}\lambda^c_{<{\bf i, i+b}>}\\
&\times&\lambda^a_{<{\bf i, i+c}>}\lambda^b_{<{\bf i, i+c}>}
\end{eqnarray}
where $a,b,c=x,\overline{x},y,\overline{y},z,\overline{z}$.
\begin{figure}
\putfig{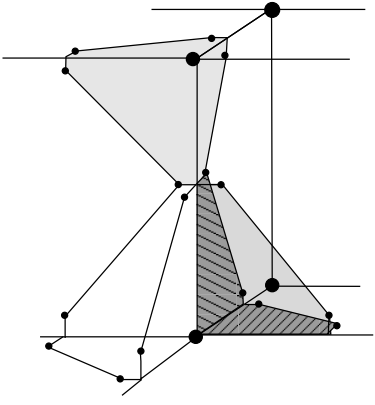}{5}
\caption{A triple of not commuting corner loops. The corner operator above does not commute with the two below. 
Each corner term affects three cubes.}
\label{3corners}
\end{figure}
The corner loop operators commute with both the link and cube operators but,
interesting enough, not all the corner loop
operators commute with the others. As it can be seen in
Fig.\ref{3corners}, a corner loop shares only one Majorana fermion with two
other corner loops belonging respectively to two other adjacent cubes. These
corner loop operators do not commute with each other when they are based on adjacent sites and the indices in
 the direction of the connecting link are conjugate and have only one index in common.
 Consider the case of corners operators at the sites ${\bf i}$ and ${\bf i+\overline{z}}$, then for example we have:
\begin{eqnarray}
\{\hat{C}^{xy\overline{z}}_{\bf i},\hat{C}^{x\overline{y}z}_{\bf i+\overline{z}}\}_+=0\\
\left[~\hat{C}^{xy\overline{z}}_{\bf i},~\hat{C}^{xyz}_{\bf i+\overline{z}}~\right]=0
\end{eqnarray}
Notice that we have the following constraint for the product of the eight corners coming from a site ${\bf i}$: $\prod_{{\bf j}=1}^8\hat{C}^{\bf j}_{\bf i}=1$. This means that on a cubic lattice with $N$ sites we have $N$ constraints on the corner operators.
 Notice also that the cube is not the product of its eight corners. Indeed, also the
product of eight corners in a cube is identically one because in the product
each Majorana fermion appears twice and they square the identity. We thus have also
the following constraint at each cube: $\prod_{{\bf j}=1}^8\hat{C}^{\bf j}=1.$
Here ${\bf j}$ of course runs instead on the eight corners belonging to the
same cube. The cube is
quite the 'square root' of $\prod_{{\bf j=1}}^8\hat{C}^{\bf j}.$ 
\begin{figure}
\putfig{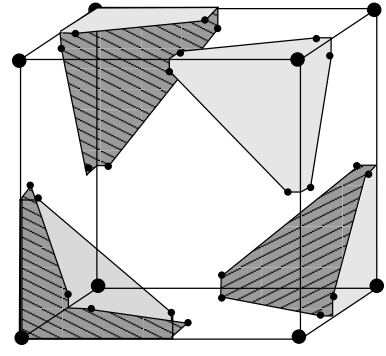}{5}
\caption{The Cube operator and four Corner Loop operators. The cube is made of
the product of only four such loops. These four operators all commute with
each other because they have no Majorana fermion in common. The four Corner
Loops shown in the figure are chosen on the odd sites.} \label{cubecorner}
\end{figure} 
Actually, the cube operator is made of the product of four of these Corner
Loop operators. We have two different ways of choosing the four corners that
make the cube out of the eight possible corners on the cube, namely the
corners on the even or odd sites, as it is shown in Fig.\ref{cubecorner}. So
we can write
\begin{eqnarray}
\sum_c \hat{S}_c=\sum_c \prod_{{\bf j\in i}(c)_{even}}\hat{C}^{\bf j}=\sum_c \prod_{{\bf j\in i}(c)_{odd}}\hat{C}^{\bf j},
\end{eqnarray}
where ${\bf j\in i}(c)_{even}$ means that ${\bf j}$ runs on the four corners
at even sites in each cube $c$,  and similarly for the odd term.

In this model we want closed membrane condensation. As we have seen, it is necessary that i) the cube operator commutes with the Hamiltonian;
ii) all the smaller pieces that compose a closed membrane must be forbidden or pay some energy, 
so that the closed membranes would not dissolve. In order to make impossible the membrane to decompose in open parts, we
can use the link term in the Hamiltonian. The border of an open membrane operator does not commute with it because they share one Majorana fermion. 
The link term has also the effect to make impossible the strings running on the links. So if this model will have closed string condensation, it will be of a new type.

\begin{figure}
\putfig{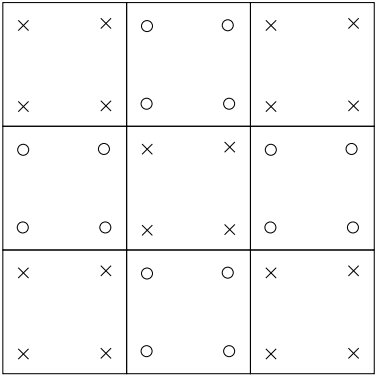}{5}
\caption{A section of the cubic lattice is shown. The picture shows the set of commuting corner operators selected for the Hamiltonian. 
At even sites, we choose the upper corners (marked with a $\circ$)
in the $xy$ and $\overline{x}\overline{y}$ quadrants and the lower corners (marked with a $\times$ in the other two quadrants. We make the 
complementary choice on odd sites. This is equivalent to alternatively take all the lower corners or all
the upper corners in a square. This means that in even cubes all the corners are taken, while in the odd ones none.
}
\label{section}
\end{figure} 

We also need to put in the Hamiltonian the corner operators. They are needed to make the model not infinite degenerate. Consider
the following set of commuting corner operators
\begin{eqnarray}
S_{even}&=&\{C_{\bf i}^{xyz},C_{\bf i}^{\overline{x}\overline{y}z},C_{\bf i}^{\overline{x}y\overline{z}},C_{\bf i}^{x\overline{yz}}\}
\\
S_{odd}&=&\{C_{\bf i}^{\overline{x}yz},C_{\bf i}^{x\overline{y}z},C_{\bf i}^{xy\overline{z}},C_{\bf i}^{\overline{xyz}}\}
\end{eqnarray}
This choice is equivalent to take all the corner operators in the even cubes and none in the odd cubes, as it is shown in fig.\ref{section}. All the operators in $S_{even}\cup S_{odd}$ commute with each other, as it is straightforward to check.
We are now ready to write down the Hamiltonian for this model. 
\be\label{fmodel}
H_{f}= -g\sum_c \hat{S}_c - h\sum_{<l>}\hat{L}_{<l>}-U\sum_{even c}\hat{C}_{\bf i}^{\bf j}
\ee
Since all the terms commute with each other, the model is exactly solvable.
The ground state manifold is
\be\label{groundstate}
\mathcal{L}=\{\ket{\psi}\in \mathcal{H}^{\otimes 3N} |  \hat{S}_c\ket{\psi}=\hat{L}_{<l>}\ket{\psi}=\hat{C}_{\bf i}^{\bf j}
\ket{\psi}=\ket{\psi}\}
\ee
What is the ground state of this model like? All the states obtained acting on a ground state with operators commuting
with the Hamiltonian are still states in the ground state. 
The ground state of the Hamiltonian (\ref{fmodel}) cannot contain open membranes, because their border does not commute
with the link term. If the membrane is closed, it is the product of cube operators and it commutes with the Hamiltonian, so 
closed membrane states are allowed in the ground state. 

It is of crucial importance to consider the non topologically trivial closed
membranes. They are non contractible closed membranes and are not the product of cubes.
A non contractible closed membrane operator is the product of all the faces on a plane $\alpha$:
\be\label{bigmembrane}
\hat{M}^{big}_{\alpha}=\prod_{s\in \alpha}\hat{F}_s
\ee 
where $\alpha=xy,yz,xz$. This big non contractible membrane still commutes with the Hamiltonian because it commutes with the cubes, the corners, and has no border
so it commutes with the links as well. The non-contractible membrane operator is of capital importance for the topological structure
of the ground state manifold, as we will see soon.

Also strings on the links are forbidden for the same reason. Small loops
corresponding to the corner loops in the Hamiltonian are allowed, but they cannot join to form bigger loops because they are disconnected.
A binary term of the type $\lambda^b_{<{\bf i, i+a}>}\lambda^c_{<{\bf i, i+a}>}$ always commutes with the links but not with the cube operators. We call it a ``hinge term''.
It does not commute with corner operators situated at an adjacent cube. This term corresponds to the edge of a cube. 
\begin{figure}
\putfig{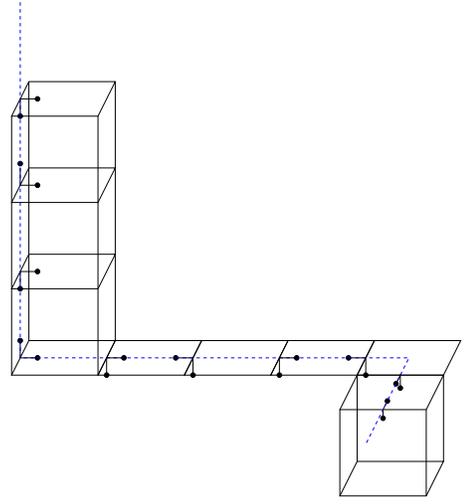}{6}
\caption{
The string featured in the model.
}
\label{bstring}
\end{figure} 
We can join the ``hinge'' terms in a string orthogonal to the links. Thus the following string operator can be defined. We first draw a string that cuts the faces in two and is orthogonal to the links. The elementary string $\gamma_a^b$ is orthogonal to the links in the $a$ direction and runs in the $b$ direction. To this elementary string we associate the operator
\be
w[\gamma_a^b]=
\lambda_{<{\bf i,i+a}>}^c\lambda_{<{\bf i,i+a}>}^{\overline{b}}
\lambda_{<{\bf i-b,i-b+a}>}^b\lambda_{<{\bf i-b,i-b+a}>}^c
\ee
where $a,b,c$ are three orthogonal directions. So each elementary string operator defines a particular cube $c$. A big string operator $\hat{W}[\Gamma]$ is the product of many elementary strings and is obtained making the product of all the $w(c)$ on the even cubes crossed by the string, see Fig.\ref{newstring}:
\be\label{newstring}
\hat{W}[\Gamma]=\prod_{even c\in\Gamma}w(c)
\ee
Because we take the ``hinge'' operators only on the even cubes, these strings always commute with the corners that we have put in the Hamiltonian. The ends of an open string do not commute with the cubes so these strings commute with the Hamiltonian only when they close. Therefore this model has a new type of closed string condensation. Notice that strings and membranes anti-commute if they intersect in a single point and thus an open string operator anti-commutes with a closed membrane operator if its end is trapped inside the membrane.

Now what about the closed membrane condensation? We established that they commute with the Hamiltonian. 
Now we have to prove that they do not dissolve in smaller pieces. 
Even cubes are elementary closed membranes that actually dissolve in the corners. But we are interested in
closed membranes of arbitrary size. Can they dissolve? A bigger closed membrane can at most ``lose'' its corners if they
belong to even cubes. Sometimes the corners get ``smoothed'' as can be seen in Fig.\ref{cubeoperator}. So closed 
membranes do not dissolve and we have a particular type of closed membrane condensation. 

Does this model have topological order? The answer is yes. It has a finite ground-state
degeneracy that is stable against perturbation. 
The degeneracy depends on the existence of a non trivial
algebra of non-contractible membranes and strings. Consider the non contractible membrane operators
(\ref{bigmembrane}) and now consider the following non-contractible string operator orthogonal to the plane $xy$:
\be\label{bigstring}
\hat{W}_{}= 
(\lambda^y_{<{\bf i, i+x}>}
\lambda^{\overline{z}}_{<{\bf i, i+x}>})
(\lambda^z_{<{\bf i-z, i-y+x}>}
\lambda^y_{<{\bf i-z, i-y+x}>})...
\ee
This non contractible string commutes with the Hamiltonian but flips all the
non contractible Membranes in the planes $xy$. These two operators realize the
four dimensional algebra $\sigma^x,\sigma^z$ on the ground state manifold
$\mathcal{L}$. We have three such algebras so the dimension of the total
algebra is $64$.  This non trivial algebra acts on the vector space
$\mathcal{L}$ which is therefore $8-$dimensional. The model has topological
order.

%%%%%%%%%%%

We can constrain this model on to a physical Hilbert space of even number of
fermions on each link by putting a constraint on the links as follows.  The constraint is
that of an even number of physical fermions on each link, so for example, on
each link $<{\bf i,i+\hat{z}}>$ we require the constraint
\begin{equation}\label{linkconstraint}
(-)^{\psi^{\dagger}_x \psi_x+\psi^{\dagger}_y\psi_y}=1
\end{equation}
of even number of fermions and analogous constraints on the links along the
other two axes. Again we can define projection operators to project down to
the physical Hilbert space. The local projection operators are obviously
\begin{equation}\label{linkprojector}
\mathcal{P}_{<{\bf i,i+\hat{a}}>}=
\frac{
1+(-)^{N_{<{\bf i,i+\hat{a}}>}}
}
{2}
\end{equation}
where
\begin{equation}
N_{<{\bf i,i+\hat{a}}>}\equiv\psi^{\dagger}_{b,<{\bf i,i+\hat{a}}>}\psi_{b,<{\bf i,i+\hat{a}}>}+\psi^{\dagger}_{c,<{\bf i,i+\hat{a}}>}\psi_{c,<{\bf i,i+\hat{a}}>}
\end{equation}
and $a\ne b\ne c$ take value in $x,y,z.$ The global projection operator is
\begin{equation}
\mathcal{P}=\prod_{links}\mathcal{P}_{<{\bf i,i+\hat{a}}>}.
\end{equation}
After the projection, the physical Hilbert space is $2^{3N}-$dimensional.
The projection makes the link term trivial: $\hat{L}_{<l>}=1$.
 This is a full local bosonic model because the total Hilbert space is a
product of finite-dimension local Hilbert spaces and the Hamiltonian is the
sum of local bosonic operators.  They are bosonic in the sense that they all
commute with each other when they are far apart.

After the projection, the model becomes a system of spins $1/2$ on the links and we can map the Hamiltonian (\ref{fmodel}) in an Hamiltonian with $\sigma$ operators acting on the links. All we have to do is to map the Corner operators correctly. The correct mapping is
\begin{eqnarray}
C_{\bf i}^{xyz}&=&\sigma^x_{<x>_{\bf i}}\sigma^x_{<y>_{\bf i}}\sigma^x_{<z>_{\bf i}}\\
C_{\bf i}^{x\overline{yz}}&=&\sigma^x_{<x>_{\bf i}}\sigma^x_{<\overline{y}>_{\bf i}}\sigma^x_{<\overline{z}>_{\bf i}}\\
C_{\bf i}^{x\overline{y}z}&=&\sigma^z_{<x>_{\bf i}}\sigma^z_{<\overline{y}>_{\bf i}}\sigma^z_{<z>_{\bf i}}\\
C_{\bf i}^{xy\overline{z}}&=&\sigma^z_{<x>_{\bf i}}\sigma^z_{<y>_{\bf i}}\sigma^z_{<\overline{z}>_{\bf i}}\\
C_{\bf i}^{\overline{x}yz}&=&\sigma^x_{<\overline{x}>_{\bf i}}\sigma^z_{<y>_{\bf i}}\sigma^z_{<z>_{\bf i}}\\
C_{\bf i}^{\overline{xyz}}&=&\sigma^x_{<\overline{x}>_{\bf i}}\sigma^z_{<\overline{y}>_{\bf i}}\sigma^z_{<\overline{z}>_{\bf i}}\\
C_{\bf i}^{\overline{xy}z}&=&\sigma^z_{<\overline{x}>_{\bf i}}\sigma^x_{<\overline{y}>_{\bf i}}\sigma^x_{<z>_{\bf i}}\\
C_{\bf i}^{\overline{x}y\overline{z}}&=&\sigma^z_{<\overline{x}>_{\bf i}}\sigma^x_{<y>_{\bf i}}\sigma^x_{<\overline{z}>_{\bf i}}
\end{eqnarray}
for the (commuting) corner operators in $\mathcal{G}$, that is, at even sites ${\bf i}$. On the odd sites ${\bf i+a}$ we have to assign the $\sigma$ operators in a complementary way, that is, sending $\sigma^x\mapsto\sigma^z$ and vice versa in order to have the right commutation-anticommutation properties.

In order to write the Hamiltonian (\ref{fmodel}) in terms of the new variables, we have to find the expression for the cube operator. It turns out that
\be
\hat{S}_c=\hat{C}^{xyz}_{\bf i}\hat{C}^{\overline{xy}z}_{\bf i+x+y}\hat{C}^{\overline{x}y\overline{z}}_{\bf i+x+z}\hat{C}^{x\overline{yz}}_{\bf i+y+z}
\ee
so we see that neighbouring cubes have complementary expressions in terms of $\sigma^x,\sigma^z$.

In terms of the $\sigma$ operators, the one-half spin model thus becomes
\begin{eqnarray}\label{1/2}
\nonumber
H_{\frac{1}{2}}= &-&g\sum_c \hat{S}_c -U\sum_{even c}\hat{C}_{\bf i}^{\bf j}\\
\nonumber
=&-&g\sum_{\bf i}
\sigma_{<x>_{\bf i}}^x \sigma_{<y>_{\bf i}}^x \sigma_{<z>_{\bf i}}^x\\
\nonumber
&\cdot&
\sigma_{<\overline{x}>_{\bf i+x+y}}^z \sigma_{<\overline{y}>_{\bf i+x+y}}^x \sigma_{<z>_{\bf i+x+y}}^x\\
\nonumber
&\cdot&\sigma_{<\overline{x}>_{\bf i+x+z}}^z \sigma_{<y>_{\bf i+x+z}}^x \sigma_{<\overline{z}>_{\bf i+x+z}}^x\\
\nonumber
&\cdot&
\sigma_{<x>_{\bf i+y+z}}^x \sigma_{<\overline{y}>_{\bf i+y+z}}^x \sigma_{<\overline{z}>_{\bf i+y+z}}^x\\
\nonumber
&-&U\sum_{even cubes}[
 \sigma_{<\overline{x}>}^z \sigma_{<y>}^x \sigma_{<z>}^z
+\sigma_{<\overline{x}>}^x \sigma_{<y>}^z \sigma_{<\overline{z}>}^x\\
&+&\sigma_{<x>}^z \sigma_{<\overline{y}>}^z \sigma_{<z>}^z 
+
\sigma_{<x>}^x \sigma_{<\overline{y}>}^x \sigma_{<\overline{z}>}^x
]
\end{eqnarray}
We can give an explicit expression for the ground states. Let $\ket{0}=\ket{0_1....0_N}$ be
the totally polarized state with all the spins down. Then if we consider the state
\be
\ket{\xi_{000}}=\prod_{c}\frac{\bbbone + \hat{S}_c}{2}\prod_s\frac{\bbbone + \hat{C}_s}{2}\ket{0}
\ee
this is obviously a ground state, indeed it is immediate to see that for any $\hat{S}_c, \hat{C}_i^j$
\be
\hat{S}_c\ket{\xi_{000}}=\hat{C}_i^j\ket{\xi_{000}}=\ket{\xi_{000}}
\ee
The cube operators generate the group of the (contractible) closed membrane operators:
\be
\mathcal{M}= \langle \hat{S}_c \rangle
\ee
the corners do not make a group because the product of corners is not a corner. Since the cubes generate a group, 
the ground state $\ket{\xi_{000}}$ contains the sum of all the possible contractible closed membrane states, immersed
in a broth of corners:
\be
\ket{\xi_{000}}=\prod_s\frac{\bbbone + \hat{C}_s}{2}\sum_{\hat{M}_{\sigma}\in\mathcal{M}}\hat{M}_{\sigma}\ket{0}
\ee
where $\hat{M}_{\sigma}$ are membrane operators defined on the contractible surfaces $\sigma$.
The other sectors of the ground state can be reached by means of the non contractible membranes $\hat{M}_{\alpha}$.
The ground state manifold can thus be written like
\be\label{gsm}
\mathcal{L}= \mbox{span}\{\ket{\xi_{ijk}}\}
\ee
where $i,j,k=0,1$ and 
\be
\ket{\xi_{ijk}}=\hat{M}_{xy}^i\hat{M}_{yz}^{j}\hat{M}_{xz}^{k}
\prod_s\frac{\bbbone + \hat{C}_s}{2}\sum_{\hat{M}_{\sigma}\in\mathcal{M}}\hat{M}_{\sigma}\ket{0}
\ee

\section{Edges of open membranes}
An open string in the model (\ref{wen3D}) has two ends that are particle-like excitations.
Open string states are excitations because they do not commute with some of
the plaquettes. Open strings have no tension, which means that their ends are
free to hop in the lattice without paying additional energy. A longer string
does not cost more energy than a shorter one. These elementary excitations
are shown to be be fermions by means of their hopping algebra \cite{lightorigin}.

In the model of Hamiltonian (\ref{1/2}), open membranes are forbidden because
their border violates the constraint on the links. So open membrane states are out of the physical Hilbert space. To make open membranes possible, we have to contour them with a ring of Majorana fermions. Then the following membrane operator defined on the open surface $\Sigma$ makes states that are within the physical Hilbert space:
\be\label{openmembrane}
\hat{M}_{\Sigma}= \prod_{s \in \Sigma}\hat{F}_s \cdot 
\prod_{{<{\bf i,i+\hat{a}}>}\in\partial\Sigma}\lambda^{a}_{<{\bf i,i+\hat{a}}>}
\ee
where $b\ne a$. Notice that $b$ can be chosen among the other three different Majorana operators that live on the link $<{\bf i,i+\hat{a}}>$, whereas the Majorana fermion $\lambda^{a}_{<{\bf i,i+\hat{a}}>}$ belongs to the face operator $\hat{F}_s$.

Now the open membrane states pay an energy because the membrane operator does not commute with the cube operators. In the model Eq.(\ref{1/2}) the elementary
excitations are open strings or open membranes. An open membrane does not
commute with the link term because of its edges. So bigger membranes cost more
energy than smaller ones: the membranes have a tension. The elementary
excitation is therefore a single edge that lives on the link of the lattice. 

The contour $\prod_{{<{\bf i,i+\hat{a}}>}\in\partial\Sigma}\lambda^{b}_{<{\bf
i,i+\hat{a}}>}$ is a closed (fermionic) string operator. This fact makes the
edge of membrane to have a fermionic property.  To give a precise definition
of a fermionic edge, let us consider a system whose linear size in $z$
direction is given by $L_z$. We also assume periodic boundary condition in the
$z$ direction and $L_z$ to be odd. Now consider an open membrane that wraps around the $z$-direction once.  The open membrane has a topology of a cylinder with two circles as its two edges.
Clearly both circles contain odd numbers of links since $L_z$ is odd.  When
the cylinder is very long, it looks like a string with two ends.  Using the
statistical algebra of Ref. \cite{levin}, we find that the ends can be viewed
as fermions in the $x$-$y$ plane.  In contrast, the edges of the membrane in
the first model (\ref{spin}) do not have such a fermionic property. It is in
this sense, we call the edge of membrane in the present model fermionic.

\section{Conclusions}
In this paper, we discussed three bosonic models on three-dimensional lattices
with non-trivial topological orders.  All models contain $Z_2$ string
condensations, and hence are described by $Z_2$ gauge theories at low
energies.  Despite this, the three model contain three different $Z_2$
topological orders. The first model (\ref{spin}) contains both string and
membrane condensations.  The ends of condensed strings (the $Z_2$ charges) and
edges of condensed membranes (the $Z_2$ vortex loops) are bosonic. The first
model gives rise to the standard $Z_2$ gauge theory at low energies.  The
second model (\ref{wen3D}) appear only to have a string condensation.  The
ends of strings are fermions.  The third model (\ref{1/2}) also contains both
string and membrane condensations. But this time, the ends of the strings are
bosonic and edges of the membranes are fermionic.  The third model gives rise
to a topological order that is not known before.

{\bf Acknowledgments}
\\
This work is partially financially supported by the European Union project
TOPQIP (Contract No. IST-2001-39215). A. H. and P.Z. gratefully
acknowledge financial support by Cambridge-MIT Institute Limited and he  Perimeter Institute of Theoretical Physics.
X. G. W. is supported by NSF Grant No. DMR--04--33632,
NSF-MRSEC Grant No. DMR--02--13282, and NFSC no. 10228408.

\end{document}